\documentclass[useAMS,usenatbib]{mn2e}

\usepackage{graphicx}
\usepackage{url}
\usepackage{subfigure}
\usepackage{latexsym}
\newcommand{\degree}{\ensuremath{^\circ}}

\newcommand{\vecg}{\bmath{g}}
\newcommand{\nhat}{\bmath{\hat n}}
\newcommand{\gradT}{\bmath{\nabla}T}

\newcommand{\WMAP}{{\sl WMAP\/}\ }
\newcommand{\Planck}{{\sl Planck\/}\ }

\title[Directionality in the \WMAP Polarization Data]{Directionality in the
\WMAP Polarization Data}
\author[D. Hanson, D. Scott and E.F. Bunn]
 {Duncan Hanson$^{1}$, Douglas Scott$^{1}$\thanks{E-mail: dscott@phas.ubc.ca}
 and Emory F. Bunn$^{2}$\\
$^{1}$Department of Physics \& Astronomy, University of British Columbia,
 Vancouver, B.C.~V6T 1Z1,~~Canada\\
$^{2}$Physics Department, University of Richmond, Richmond, VA 23173,~~USA}
\begin{document}
\setcounter{figure}{0}
\bibliographystyle{mn2e}

\date{Accepted 2007 July 3. Received 2007 June 27; in original form 2007 February 9}
\pagerange{\pageref{firstpage}--\pageref{lastpage}} \pubyear{2007}
\maketitle

\label{firstpage}

\begin{abstract}
Polarization is the next frontier of CMB analysis, but its signal is dominated
over much of the sky by foregrounds which must be carefully removed. To
determine the efficacy of this cleaning it is necessary to have sensitive
tests for residual foreground contamination in polarization sky maps. The
dominant Galactic foregrounds introduce a large-scale anisotropy on to the
sky, so it makes sense to use a statistic sensitive to overall directionality 
for this purpose.  Here we adapt the rapidly computable $\cal{D}$ statistic of 
Bunn and Scott to polarization data, and demonstrate its utility as a foreground 
monitor by applying it to the low resolution \WMAP 3-yr sky maps. With a 
thorough simulation of the maps' noise properties, we find no evidence for 
contamination in the foreground cleaned sky maps.
\end{abstract}


\section{Introduction}

Our progress in understanding the large scale make-up of the Universe has been
propelled rapidly in the last 15 years by extremely detailed measurements of
the Cosmic Microwave Background (CMB). Current measurements of the CMB
temperature anisotropies have allowed us to place stringent constraints on
some of our Universe's key parameters and are proving to be a driving force
in fundamental physics (see e.g.~\citealt{ScoSmo}).
Cosmic-variance places limits on the amount of
information which can be extracted from the temperature anisotropies alone,
however. The \Planck satellite, for example, is expected to reach this limit
over the entire primary temperature anisotropy power spectrum \citep{Blue}. Fortunately, the polarization of the CMB anisotropies provides us with a
wealth of additional cosmological information which we have only just begun
to probe. The benefits of CMB polarization data are numerous. Perhaps most
excitingly, polarization data provide a chance to detect primordial
gravitational waves, through the measurement of B-mode polarization
\citep{SelZal,KamKosSteb,HuWhi,KamKos}. The rapid improvements in CMB
polarization data provided by experiments like PIQUE \citep{Hed},
DASI \citep{Kov}, CAPMAP \citep{Bark}, BOOMERanG \citep{Masi},
CBI \citep{Read}, MAXIPOL \citep{Wu}, \WMAP \citep{Page}, and  upcoming
experiments like BICEP, CLOVER \citep{Tay}, QUAD, and the \Planck satellite,
ensure that polarization will become one of the most important tools we have
for furthering our understanding of the Universe in the years to come.

Unfortunately, the polarization signal is much more difficult to measure than
the temperature anisotropies. The magnitude of the CMB polarization
anisotropies is at most 10 per cent that of the temperature anisotropies, making
their detection alone a challenging task. This difficulty is being gradually
overcome with improvements in detector technology. Another
problem with the CMB polarization, however, is that its signal is dominated
over most of the sky by foregrounds. In particular, over the wavelength range
where the CMB is typically measured, synchrotron and non-spherical dust
emission from within our Galaxy contaminate uncleaned polarization maps. It is
not yet clear how well the problem of foreground contamination can be solved.
Its resolution will require both a detailed knowledge of the structure and
magnitude of the foregrounds, and an understanding of how to remove them from
the polarization data without introducing unwanted systematics.

Understandably, the issue of polarization foregrounds has received much
attention in the literature. Several papers have been written on the
construction of templates for the emission \citep{deOliv,Bacc,Page,Hans}, and
a number of methods have been proposed for performing the foreground removal
\citep{BouPruSet,Hans,SloSelMak}.  Testing the efficacy of these methods is a
crucial step in the analysis pipeline, to compare the abilities of the various
removal techniques and ultimately to provide confidence in the accuracy of CMB
polarization data. Tests for the existence of residual polarized foreground
contamination have received less attention in the literature, however. The
test used in the \WMAP team's analysis, where foregrounds are cleaned using a
template method in pixel space, is a simple ${\chi}^2$ statistic \citep{Page}.
Unfortunately, this statistic cannot be used to make comparisons between
different cleaning methods. The only currently proposed `general' statistic which is
capable of this application is based on the Bipolar Power Spectrum (BiPS)
and has been developed by \citet{BasHajSou}.

In this paper, we introduce and analyse another general method of testing for
residual foreground contamination in CMB polarization maps, the $\cal{D}$
statistic of \citet{BunSco}. It possesses the virtues of having a simple
interpretation as a measure of directionality in a CMB map and of being
extremely rapid to compute. We argue that this statistic is well suited to
the detection of foreground contamination, as both the synchrotron and dust
foreground components have strong overall directionality on large scales. We
then apply the $\cal{D}$ statistic to the \WMAP 3-yr polarization maps to
demonstrate its usefulness for real data.

\section{The $\cal D$ Statistic}
\label{section.thedstatistic}
The $\cal D$ statistic was first presented by \citet{BunSco}; the $\cal D$
value for a sky map pixelized into $N$ pixels is defined as
\begin{equation}
{\cal D}\equiv {\max_{\nhat} f(\nhat)\over\min_{\nhat} f(\nhat)},
\label{eq.boneheaddef}
\end{equation}
where $\nhat$ runs over the unit sphere, and with the function $f(\nhat)$
given by 
\begin{equation}
f(\nhat)=\sum_{p=1}^{N} w_p (\nhat \cdot \vecg_p)^2.
\label{eq.fdef}
\end{equation}
Here $\it{w_p}$ are weights chosen to compensate for noise and the effects of
a cut sky, and $\vecg_p$ is a local directionality vector calculated for the
pixel $p$. The $f(\nhat)$ value is a measure of how much (or how little) the
$\vecg_p$ of a map tend to point in a given direction. The $\cal D$ statistic
tells us how extreme these values are. Once $\cal D$ has been calculated for
real sky data, we can compare its value to that found for simulated
realizations of the CMB. Calculating $\cal D$ for a large number of these
simulations gives us a histogram of expected values, and excess directionality
in a CMB data-set appears as a value of $\cal D$ which is an outlier of this
distribution.

For sky maps of temperature, a reasonable quantity to use for
$\vecg_p$ is the local temperature gradient $\gradT_p$. For polarization maps,
on the other hand, the polarization psuedo-vectors
themselves are obvious candidates. We define $\vecg_p$
to be a vector whose magnitude and direction are those of the polarization
at point $p$.  To express $\vecg_p$ explicitly in terms of the
Stokes parameters $Q$ and $U$, recall that the polarization magnitude
is
\begin{equation}
P = \sqrt{{Q}^2 + {U}^2},
\label{eq.vecgmag}
\end{equation}
while the polarization direction lies in the tangent plane to the celestial
sphere at $p$ and makes an angle with the meridian of
\begin{equation}
\theta = \frac{1}{2}\tan^{-1}\left(\frac{U}{Q}\right).
\label{eq.vecgtheta}
\end{equation}
Measuring the angle from the meridian is a \WMAP\ convention \citep{Page}
which we adopt here, although it is not universally followed.  We define the 
vector $\vecg_p$ to have magnitude $P$ and direction given by $\theta$.

Because the polarization is a spin-2 quantity, it is represented
by headless `pseudo-vectors' rather than vectors, so $\theta$
can always be rotated by $180^\circ$ without altering the polarization.
However, due to the quadratic definition of $f(\nhat)$, this ambiguity
does not affect $\cal{D}$.  Moreover,
because the function $f$ smoothly averages together polarization
information over the whole sky,
$\cal D$ is only sensitive to large scale ($\ga20\degree$)
directionality.

The weights $w_p$ must be chosen to satisfy the null hypothesis that for
isotropically distributed $\vecg_p$ no preferred direction will be found on
average, even when parts of the sky have been removed. We can see how these
weights may be obtained by casting $f(\nhat)$ in matrix form as
\begin{equation}
f(\nhat)=\nhat^T \bmath{A}\, \nhat,
\label{eq.fmatrixdef}
\end{equation}
where $\bmath{A}$ is a $3\times3$ matrix defined by
\begin{equation}
A_{ij} = \sum_{p=1}^{N} \it{w_p g_{pi} g_{pj}}.
\label{eq.amatrixdef}
\end{equation}

In order for no preferred direction to be found, we must have $f(\nhat)$
independent of $\nhat$ (on average). From equation~\ref{eq.fmatrixdef} it can be seen that
this constraint may be realized by choosing suitably normalized $w_p$ such
that $\langle \bmath{A} \rangle$ is equal to the identity matrix for isotropically distributed
$\vecg_p$. This criterion ensures that $f(\nhat)$ will be approximately
constant over the sky when the underlying data lack a preferred direction,
although it does not guarantee that fluctuations of $f$ about its ensemble
average will be isotropic. As a result, $\cal D$ may still pick out preferred
directions arising from the noise structure or sky coverage, even when the
underlying signal is isotropic. We find that such departures from anisotropy
are weak in practice; this point is discussed further in
Section~\ref{section.wmappolarizationanalysis}.

Because $\bmath{A}$ is symmetric, requiring $\langle \bmath{A} \rangle = \bmath{I}$ gives only
6 independent equations on the weights. To completely determine them for a
large number of pixels, the additional constraint is imposed that
${\rm Var}(\it{w_p}P_p)$ be minimized, where $P_p$ is the expected value of
$|\vecg_p|^2$ given by pixel noise and cosmological signal. For a large
signal-to-noise ratio on an isotropic sky this is equivalent to minimizing
the variance of the weights. When noise is not negligible, minimizing
${\rm Var}(\it{w_p}P_p)$ rather than ${\rm Var}(\it{w_p})$ has the favourable
effect of down-weighting noisy pixels. More explicit details on the
calculation of the weights are given in \cite{BunSco}.

Once the weights $w_p$ have been calculated, the simple quadratic definition
of $f(\nhat)$ makes it possible to calculate $\cal D$ for a given data-set
in $O(N)$ operations, where $N$ is the number of pixels. This can be done by
noting that the values of $\nhat$ which extremize $f(\nhat)$ are given by the
eigenvectors of $\bmath{A}$.  $\cal{D}$ follows immediately by taking the
ratio of the largest and smallest eigenvalues.

$\cal D$ is one of the simplest statistics which can be proposed for the identification of statistical anisotropy in a CMB map. To compare it to the BiPS method of \cite{BasHajSou}, we note that it is ultimately a less powerful 
statistic, as BiPS can be used to probe both large and small scales, whereas 
$\cal D$ is only sensitive to large scale anisotropies. The advantage of $\cal{D}$,
however, is that it can be calculated in $O(N)$ operations,
an unbeatable scaling. This speed makes $\cal D$ an excellent statistic for
`first look' determination of anisotropy in CMB maps, and the concomitant
possible foreground contamination.

\section{Directionality of polarization foregrounds}
\label{section.directionalityofpolarizationforegrounds}
The two main sources of foreground contamination in CMB polarization maps at
the ${\sim}\,100\,$GHz frequencies where the CMB is usually probed are
Galactic synchrotron and thermal dust emission.
Polarized synchrotron emission results from the acceleration of cosmic-ray
electrons as they orbit in the Galactic magnetic field. Polarized dust
emission, on the other hand, is the result of non-spherical dust grains which
tend to align their long axes perpendicular to the field; these dust grains
preferentially emit thermal radiation polarized along their long axis
\citep{DavGre}.

The polarization of both the synchrotron and thermal dust emission has its
origins in the anisotropy introduced by the Galactic magnetic field, and at
CMB wavelengths (where Galactic Faraday rotation is negligible) these two
components should both be polarized preferentially in the same direction.
Thus, we expect that they have a common and distinctive directionality on the
sky, dictated by the large scale structure of the Galactic magnetic field. To
illustrate the signature of this directionality under application of the
$\cal{D}$ statistic, we analyse the low resolution K-band polarization map at
$23\,$GHz and the dust polarization template of the \WMAP team, masking both
with the standard polarization mask (P06) to mimic the situation when analysing
maps for foreground contamination. For the K-band map we find a preferred
direction at a Galactic latitude $86.4\degree$, within $4\degree$ of the
poles. For the dust emission template we find that the preferred direction
lies at $79.1\degree$, $11\degree$ shy of the poles. Thus, the
significant presence of synchrotron and dust foregrounds in a polarization
map is expected to result in an outlying value of $\cal{D}$, with a preferred
direction in the vicinity of the Galactic poles.

\section{Analysis of \WMAP 3-yr polarization data}
\label{section.wmappolarizationanalysis}
The release of the \WMAP 3-year data has provided us with a glimpse at the
structure of CMB polarization.
Here we analyse the $Q$ and $U$ polarization maps for foreground contamination
using the $\cal D$ statistic.

We have chosen to examine the \WMAP low resolution 3rd year (v2) polarization
maps\footnote{Available at \url{http://lambda.gsfc.nasa.gov/}} at
{\sc HEALPix}\footnote{See \url{http://healpix.jpl.nasa.gov}} resolution 4,
corresponding to 3072 pixels on the entire sphere. These are the maps used
for low $\ell$ analysis of the \WMAP data, corresponding to the angular scales which
$\cal D$ effectively probes. At this resolution it is also feasible to
accurately simulate realizations of the \WMAP\ noise structure using the full
noise covariance matrix, which becomes necessary for large angle analyses.

\begin{figure}
\begin{center}
\includegraphics[width=8.5truecm]{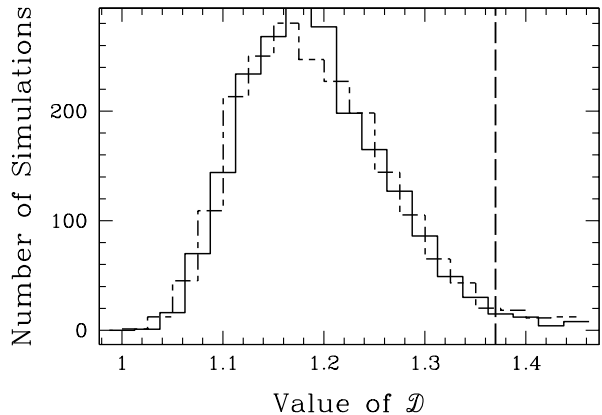}
\caption{Directionality histograms for maps simulated with \WMAP low
resolution V-band noise levels, with (solid) and without (dot-dashed)
cosmological signal added. The two are essentially indistinguishable. The
dashed line shown at ${\cal D} = 1.37$ is the value of $\cal D$ calculated
for the V-band data themselves.}
\label{fig.dhistforVpolwwocosmo}
\end{center}
\end{figure}

In all of our simulations, we generate cosmological signals given by the
`\WMAP\hspace{-.05in}+all $\Lambda$CDM' fit parameters \citep{Sper}. The details of this choice have
little effect on the outcome of our simulations. All of the polarization maps
are strongly noise-dominated, and we find that histograms of $\cal D$ for
simulated data with and without cosmological signal are indistinguishable for
all of the \WMAP bands; this is illustrated in
Fig.~\ref{fig.dhistforVpolwwocosmo} for simulated V-band maps. Thus, we
believe that, with an accurate treatment of the maps' noise properties, any
excess directionality which we detect in the \WMAP data can be attributed to
residual foregrounds, and not a cosmological signal.

For all of our simulations, we add correlated Gaussian noise to each unmasked map pixel based on the \textit{full QU covariance matrices} for the corresponding \WMAP frequency band, as provided by the \WMAP team. We generate this noise using the Cholesky decomposition of the covariance matrix. This method accounts for QQ, QU and UU correlations both within each pixel and with other pixels across the sky.

\begin{figure}
\begin{center}
\includegraphics[width=8.5truecm]{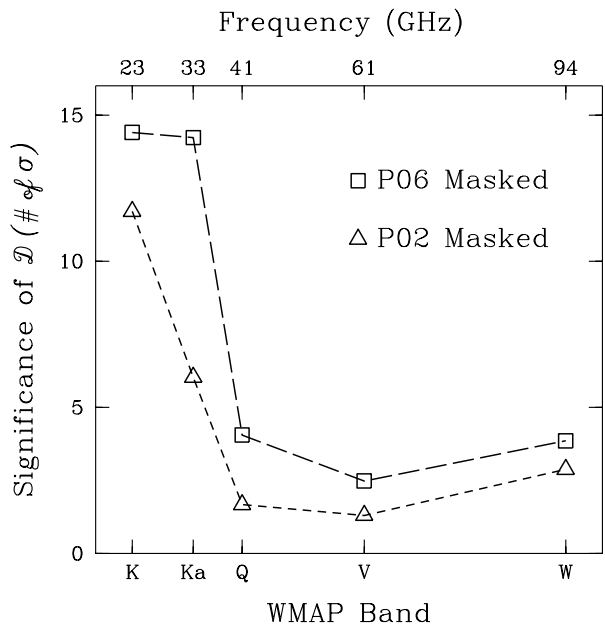}
\caption{Significances of foreground contamination for the \WMAP low resolution
maps using noise based on the full covariance matrix. The ($\Box$) symbols are for the standard P06 mask (27 per cent sky cut),
while the ($\bigtriangleup$) symbols are for the more stringent P02 masking
(50 per cent sky cut for our downgrading scheme). The curve roughly follows that expected from the
frequency dependence of contamination by synchrotron and dust foreground
components.}
\label{fig.significancesforintrapixel}
\end{center}
\end{figure}

For uncleaned maps, the $\cal D$ statistic produces the expected results. For
each band, we generate and analyze ${\sim}\,2000$ simulated skies. An example of this is shown in
Fig.~\ref{fig.dhistforVpolwwocosmo}. We can use the number of standard
deviations (at which the real data lie from the histogram mean) as an
approximate measure of the significance with which we detect foreground
contamination. This is illustrated in Fig.~\ref{fig.significancesforintrapixel},
which shows the expected dependence on radiometer band as the synchrotron
emission decreases with frequency and the dust contribution increases,
leading to an apparent minimum of foreground contamination near the V-band. We can see
that use of the standard P06 polar mask results in a greater level of
contamination than when we use the more stringent P02 mask.
For the P06 mask, all of the preferred directions which are found for the
real data lie within $10\degree$ of the Galactic poles, and for the P02 mask
the preferred directions all lie within $25\degree$ of the poles, further
demonstrating the foreground origin of the directionality excess. With P06 masking, we find that $98\%$ of our V-band simulations have values of $\cal D$ less than the value calculated for the \WMAP data. With P02 masking, this result drops to $90\%$. The corresponding significance figures of merit are $2.5\sigma$ and $1.3\sigma$ respectively. 

\begin{figure}
\begin{center}
\includegraphics[width=8.5truecm]{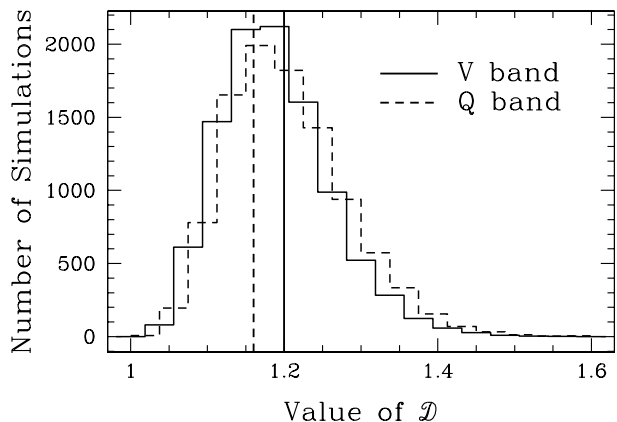}
\caption{Directionality histograms for simulated realizations of the \WMAP
V-band (solid) and Q-band (dashed) `low resolution foreground reduced' maps,
using the full covariance matrix description of the maps' noise properties.
The vertical lines are the values of $\cal D$ calculated for the Q- and V-band
data themselves.}
\label{fig.QVforeredhists}
\end{center}
\end{figure}

In order to assess the efficacy of the foreground removal performed by 
the \WMAP team we also analyse the low resolution Q- and V-band foreground
reduced maps. These have had dust and synchrotron templates projected out to
remove foreground contamination. The results of these analyses are shown in Fig.~\ref{fig.QVforeredhists}.
From this figure it can be seen that the $Q$ and $V$ maps have similar noise
properties under analysis with $\cal D$, as the simulation histograms almost
completely overlap.  We find no evidence for residual foreground contamination
in these maps, their $\cal D$ statistics both lying well within the range of
probable values, and with the maximal $\cal D$ directions for the
Q- and V-band data being $55\degree$ and $27\degree$ from the pole,
respectively. Our significance figure of merit is $0.6\sigma$ for the Q-band and $0.16\sigma$ for the V-band maps. $31\%$ of the simulated $\cal D$ values lie below the map value for the Q-band, and $60\%$ of the simulated data lie below the V-band value. Thus, we find no evidence for foreground contamination in these cleaned maps.

\begin{figure}
 \begin{center}
    \mbox{
      \subfigure[Diagonal]{\scalebox{0.3}{\includegraphics{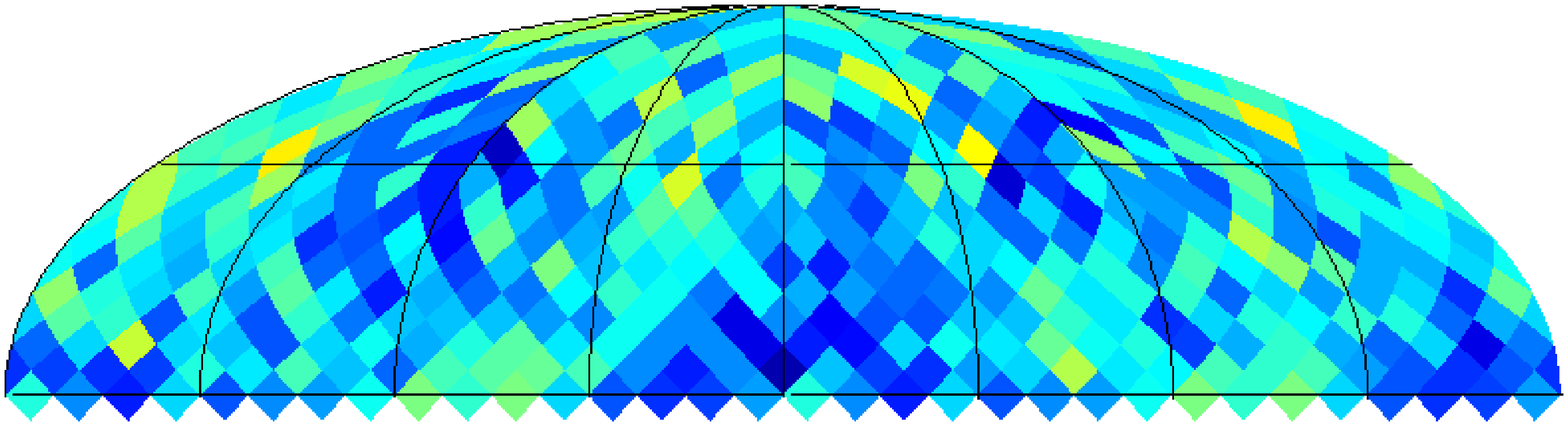}}}
      }
      \mbox{
       \subfigure[Intrapixel]{\scalebox{0.3}{\includegraphics{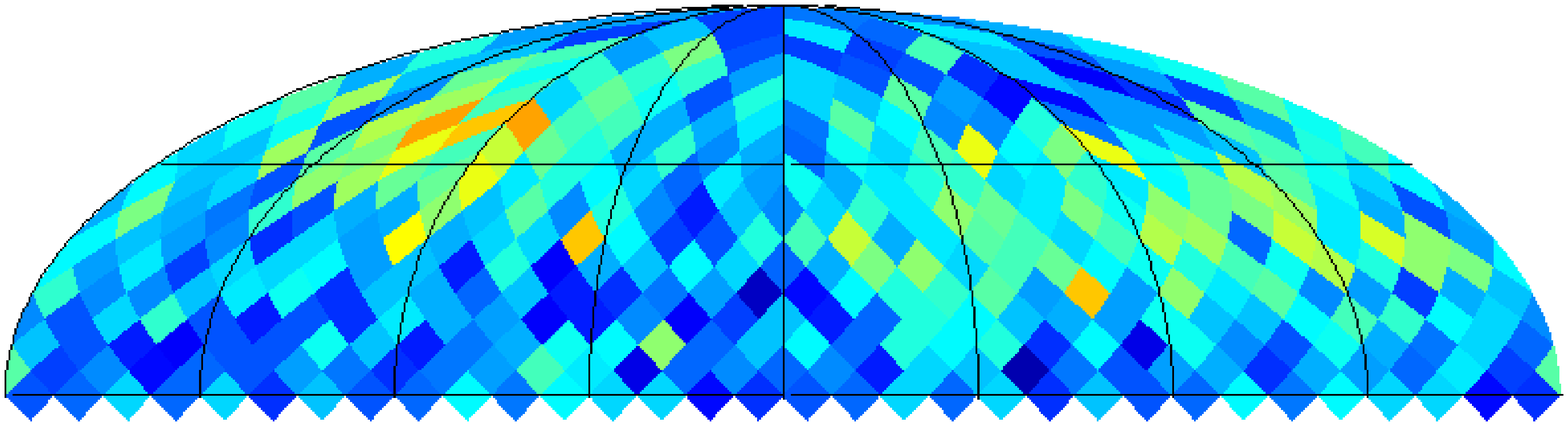}}} 
       }
      \mbox{
       \subfigure[Full covariance matrix]{\scalebox{0.3}{\includegraphics{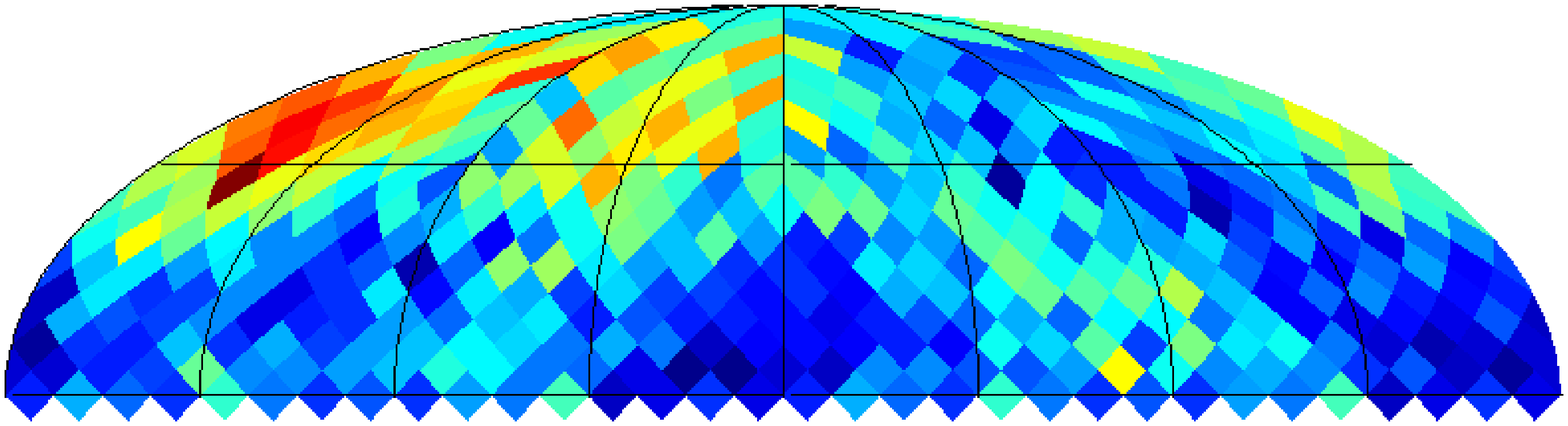}}} 
       }
       \mbox{
       	\includegraphics[width=6.5truecm]{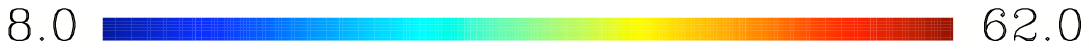}
        }
    \caption{Histograms of preferred direction found for maps simulated with
diagonal, intra-pixel, and full covariance matrix based noise (Mollweide
projection, Galactic coordinates). The colour units are in number of
simulations.  The preferred directions calculated by $\cal D$ are headless,
and so we have taken the convention of choosing them to be in the Northern
Hemisphere. }
    \label{fig.diagintrafullcovhists}
  \end{center}
\end{figure}

It is interesting to note that if we had generated noise which only accounted
for the intra-pixel or diagonal components of the noise covariance matrix then we would
have found strong evidence for residual contamination. In the case of the V-band foreground reduced data, for example, we would have found some evidence for foreground contamination, with over $98\%$ of simulated $\cal D$ values falling below that calculated for real data in the cleaned V-band maps (corresponding to a value of $2.5\sigma$ for our significance figure). Accounting for \textit{inter}-pixel noise correlations broadens the distribution of
simulated $\cal D$ values. Indeed, the detailed noise structure of the \WMAP
data appears to be such that it introduces a degree of anisotropy into the
maps. This can best be seen by looking at the histograms of preferred
directions found in a series of 30,000 V-band map simulations, given in
Fig.~\ref{fig.diagintrafullcovhists}. Differing methods of noise generation
were utilized for three sets of simulations. `Diagonal' noise was generated
from only the $QQ$ and $UU$ elements of the covariance matrix, intra-pixel
noise included $QU$ terms, and the full covariance matrix treatment simulated
inter-pixel correlations as well. Unlike the isotropic distribution of
preferred directions which is found when accounting only for diagonal noise
correlations, there is a tendency towards anisotropy for the intra-pixel and
full covariance simulations, with a definite slight preference for directions
in the vicinity of $(l,b) = (150\degree, 40\degree)$ for the full covariance
simulations.  The statistics of these two dimensional histograms support this
visual finding, with the diagonal, intrapixel, and full covariance simulation
methods having variance/mean values of $1.41$, $1.66$ and $3.80$ respectively.
For an isotropic sky, we expect the histogram values to follow a multinomial
distribution, and to have a variance/mean of $1 \pm .08$. We believe that the
excess variance which is seen in the diagonal and intra-pixel simulation histograms can be
attributed to the slight inherent bias possessed by $\cal D$, as discussed in
Section~\ref{section.thedstatistic}. The much higher excess variance seen in
the inter-pixel simulations can be attributed to the sum of this bias and another, 
stronger bias given by the noise covariance structure of the map. This bias is independent of the
specifics of the $\cal D$ statistic, and most likely can be detected in the future by other
polarization statistics sensitive to anisotropy. The dependence of our results on the precise treatment of noise  
underscores the general need for care when dealing with large angle polarization data.

\section{Conclusion}
We have investigated the $\cal D$ statistic of \citet{BunSco} in the context
of CMB polarization data, finding that it is an excellent general tool for
quickly assessing the magnitude of foreground contamination in polarization
maps. Under analysis with $\cal D$, imperfectly removed foregrounds leave a
tell-tale signature of excess directionality toward the Galactic poles.  An
analysis of the \WMAP 3-year polarization data supports these statements. We
find the expected dependence of foreground contamination on channel frequency
in all of the \WMAP low resolution maps, with the magnitude of this excess
disappearing in the cleaned, `foreground reduced' maps.

In the future, a better understanding of the polarization foregrounds will
allow us to develop more sophisticated statistics to precisely evaluate the
possibility of foreground contamination in cleaned maps. Given the current
state of understanding, however, we feel that calculation of $\cal D$ is a
useful technique for this purpose, as it is one of the simplest possible
statistics which captures the directional signature of the polarization
foregrounds. It has the advantages of being simple and rapidly computable,
and can be considered complementary to other statistics such as the BiPS
\citep{BasHajSou}.  Hence it should be one of the diagnostic tests ready to
assess foreground contamination in data from the \Planck satellite, as well
as other large-angle polarization experiments.

\section{Acknowledgments}
The work carried out in this paper made use of the {\sc HEALPix} \citep{Gorsk} package
for pixelization.
This research was supported by the Natural Sciences and Engineering Research
Council of Canada and the Canadian Space Agency.
We also acknowledge the contributions of Patrick Plettner for some related
earlier work.

\label{lastpage}
\end{document}